\documentclass[aps,pra,twocolumn,letterpaper,longbibliography,superscriptaddress,floatfix,amssymb,amsmath]{revtex4-1}

\usepackage[colorlinks,urlcolor=black,citecolor=black,linkcolor=black]{hyperref}
\usepackage{graphicx}
\usepackage{amsmath,amssymb,bm}
\usepackage{xcolor}

\newcommand{\micron}{\ensuremath{\mu\mathrm{m}}}

\newcommand{\threevec}[3]{\left(\begin{array}{c}#1\\#2\\#3\end{array}\right)}

\newcommand{\bvec}[1]{\mathbf{#1}}

\newcommand{\inleva}[1]{\langle#1\rangle}

\newcommand{\SO}{\ensuremath{\mathrm{SO}}}

\newcommand{\U}{\ensuremath{\mathrm{U}}}

\newcommand{\nematic}{\mathbf{\hat{d}}}
\newcommand{\absF}{\ensuremath{|\langle\bvec{\hat{F}}\rangle|}}

\newcommand{\xhat}{\ensuremath{\mathbf{\hat{x}}}}
\newcommand{\yhat}{\ensuremath{\mathbf{\hat{y}}}}
\newcommand{\zhat}{\ensuremath{\mathbf{\hat{z}}}}

\newcommand{\rr}{\ensuremath{\mathbf{r}}}

\newcommand{\nuke}[1]{}

\newcommand{\beq}{\begin{equation}}
\newcommand{\eeq}{\end{equation}}

\begin{document}

\title{Controlled Creation and Decay of Singly-Quantized Vortices \\ in a Polar Magnetic Phase}

\author{Y.~Xiao}
\altaffiliation[Current Address: ]{Department of Electrical and Computer Engineering, University of Michigan, Ann Arbor, Michigan 48109, USA}
\affiliation{Department of Physics and Astronomy, Amherst College, Amherst, Massachusetts 01002--5000, USA}

\author{M.~O.~Borgh} 
\affiliation{Faculty of Science, University of East Anglia, Norwich, NR4 7TJ, United Kingdom}

\author{L.~S.~Weiss}
\altaffiliation[Current Address: ]{Department of Physics and James Franck Institute, University of Chicago, Chicago, Illinois
60637, USA}
\affiliation{Department of Physics and Astronomy, Amherst College, Amherst, Massachusetts 01002--5000, USA}

\author{A.~Blinova}
\affiliation{Department of Physics and Astronomy, Amherst College, Amherst, Massachusetts 01002--5000, USA}
\affiliation{Department of Physics, University of Massachusetts Amherst, Amherst, Massachusetts 01003, USA}

\author{J.~Ruostekoski} 
\affiliation{Department of Physics, Lancaster University, Lancaster, LA1 4YB, United Kingdom}

\author{D.~S.~Hall}
\affiliation{Department of Physics and Astronomy, Amherst College, Amherst, Massachusetts 01002--5000, USA}

\date{\today}

\begin{abstract}
We experimentally and theoretically explore the creation and time evolution of vortex lines in the polar magnetic phase of a trapped spin-1 $^{87}$Rb Bose--Einstein condensate. A process of phase-imprinting a nonsingular vortex, its decay into a pair of singular spinor vortices, and a rapid exchange of magnetic phases creates a pair of three-dimensional, singular singly-quantized vortex lines with core regions that are filled with atoms in the ferromagnetic phase. Atomic interactions guide the subsequent vortex dynamics, leading to core structures that suggest the decay of the singly-quantized vortices into half-quantum vortices.
\end{abstract}


\maketitle

Vortices in superfluids with internal degrees of freedom, such as those existing within spinor Bose--Einstein condensates (BECs)~\cite{kawaguchi_physrep_2012,stamper-kurn_rmp_2013} and superfluid liquid He-3~\cite{volovik,vollhardt-wolfle}, exhibit a much richer phenomenology than do simple line vortices in scalar superfluids. Notable examples abound, including vortices with fractional charges~\cite{leonhardt_jetplett_2000,ruostekoski_prl_2003,ji_prl_2008,lovegrove_pra_2012,seo_prl_2015,autti_prl_2016,kang_prl_2019}, vortices with like charges that sum to zero~\cite{weiss_ncomm_2019}, vortices with charges that do not commute~\cite{kobayashi_prl_2009,borgh_prl_2016,mawson_pra_2015,semenoff_prl_2007,barnett_pra_2007,borgh17,mawson_prl_2019}, and nonsingular textures with angular momentum~\cite{mizushima_prl_2002,martikainen_pra_2002,leanhardt_prl_2003,leslie_prl_2009,choi_njp_2012,lovegrove_prl_2014}. These features, \emph{inter alia,} hint at their highly counter-intuitive dynamics.

The symmetry properties of the superfluid order parameter determine its magnetic phases and topologically permissible vortex excitations~\cite{kawaguchi_physrep_2012,supplemental}. The ground state of a spin-1 system, for example, exhibits two phases: a polar phase, which minimizes the total spin and is characterized by a nematic axis $\nematic$ and condensate phase $\tau$; and a ferromagnetic (FM) phase, which maximizes the total spin and is characterized by a vector triad. In turn, the ground-state phase of an atomic BEC at zero magnetic field is determined by the nature of the interatomic interactions, which are themselves polar (e.g., in $^{23}$Na) or FM (e.g., in $^{87}$Rb)~\cite{kawaguchi_physrep_2012}. Thus do interactions at the atomic scale influence both the type and destiny of vortices within the condensate.

In the polar phase, a singly-quantized vortex (SQV) with $2\pi$ phase winding is unstable against splitting into a pair of half-quantum vortices (HQVs), each with $\pi$ phase winding. This unusual possibility was proposed and analyzed in Ref.~\cite{lovegrove_pra_2012} and subsequently observed experimentally within a $^{23}$Na BEC in an effectively two-dimensional trapping geometry~\cite{seo_prl_2015,seo_prl_2016}. The core of a vortex in superfluid $^3$He-$B$ has similarly been predicted~\cite{thuneberg_prl_1986,salomaa_prl_1986} and observed~\cite{kondo_prl_1991} to consist of two HQVs; and, more recently, HQVs have been observed in the $^3$He polar phase~\cite{autti_prl_2016}.

In this Letter we describe the controlled creation and subsequent time-evolution of a pair of three-dimensional (3D) singular SQVs in the polar phase of a spin-1 $^{87}$Rb BEC with FM interatomic interactions. In contrast to techniques that randomly nucleate vortices throughout the superfluid by, e.g., stirring~\cite{rosenbusch_prl_2002,neely_prl_2010,seo_prl_2015} or rapid cooling through the superfluid transition~\cite{tilley-tilley,weiler_nature_2008,freilich_science_2010}, our experiment makes use of a deliberately applied strong bending of a nonsingular spin texture to generate a single pair of $\SO(3)$ vortices with polar cores at a specific location within the BEC~\cite{weiss_ncomm_2019}. A sudden exchange of the polar and FM phases results in the desired pair of polar SQVs, where the topological interface~\cite{borgh_prl_2012,lovegrove_pra_2016} between the two magnetic phases within each vortex core is imaged directly. In a final step, a radio-frequency $\pi/2$ spinor rotation causes each SQV to evolve towards a pair of HQVs. We numerically model these experimental conditions and show how the FM interactions influence and complicate the decay process as compared with polar interactions.

The theoretical analysis uses the mean-field model for a spin-1 BEC, with Hamiltonian density~\cite{kawaguchi_physrep_2012,stamper-kurn_rmp_2013}
\begin{equation}
\label{eq:hamiltonian}
\mathcal{H} = h_0 + \frac{c_0}{2}n^2 + \frac{c_2}{2}n^2|\mathbf{\inleva{\hat{F}}}|^2 - pn\inleva{\hat{F}_z}+qn\inleva{\hat{F}_z^2}
\end{equation}
for the spinor wavefunction
\begin{equation}
  \label{eq:spinor_suppl}
  \Psi({\bf r}) = \sqrt{n({\bf r})}\zeta({\bf r})
  = \sqrt{n({\bf r})}\threevec{\zeta_+({\bf r})}
                              {\zeta_0({\bf r})}
                              {\zeta_{-}({\bf r})},
  \quad
  \zeta^\dagger\zeta=1
\end{equation}
expressed in a basis quantized along the $z$ axis. Here, $\mathbf{\hat{F}}$ is the vector of spin-1 matrices, $h_0 = \hbar^2/(2M)|\nabla \Psi|^2 + (M\omega_r^2/2)(x^2+y^2+2z^2)n$ for atomic mass $M$ and radial trap frequency $\omega_r$, and $n=\Psi^\dagger\Psi$ is the atomic density. The constants $c_0$ and $c_2$ parameterize the spin-independent and spin-dependent interaction strengths, respectively, and $p\propto |\mathbf{B}|$ and $q \propto |\mathbf{B}|^2$ give the linear and quadratic Zeeman energy shifts due to an applied magnetic field $\mathbf{B}$.

The ground state of the system is determined by the sign of the interaction strength $c_2$ and, at fixed magnetization, the quadratic Zeeman term $q$ \cite{supplemental}. In our experiment $q > |c_2| n$, specifying an easy-axis polar (EAP) ground-state phase~\cite{kawaguchi_physrep_2012}. We shall see that this introduces significant dynamics when $\nematic$ is not aligned with the magnetic field.

The experiment begins with an FM $^{87}$Rb condensate of $N\sim 2.0 \times 10^5$~atoms in an optical trap with frequencies $(\omega_r,\omega_z) = 2\pi(130,170)~\mathrm{s}^{-1}$. The atoms are exposed to a magnetic field described by
\begin{align}
\mathbf{B}(t) = B_z(t)\zhat + (\xhat+\yhat-2\zhat)b_q(t),
\end{align}
where $B_z$ is the strength of an applied bias field along the $z$ axis and $b_q$ is the strength of a 3D quadrupole field produced by a pair of anti-Helmholtz coils. The condensate spin is initially aligned along the $z$ axis, represented by the spinor $(1,0,0)^\mathrm{T}$ in the space-fixed basis we adopt for the remaining discussion.

We use a phase imprinting process to introduce the polar SQVs, initially following the vortex creation technique introduced in Ref.~\cite{weiss_ncomm_2019}. A nonsingular vortex is first created by linearly ramping the magnetic bias field $B_z$ from $0.03$~G to $-0.05$~G at $-5$~G/s, with $b_q(0)=4.3(4)$~G/cm. The atomic spins incompletely follow the nonadiabatic reorientation of the magnetic field~\cite{NAKAHARA00,pietila_prl_2009_dirac} as its zero passes through the condensate, resulting in the desired spin texture. Immediately afterwards we ramp the field to its minimum value $-0.38$~G in 10~ms, eliminate the magnetic quadrupole contribution $b_q \rightarrow 0$, and adiabatically reorient the field to 1~G along the $+z$ axis. In the subsequent 100~ms, the tight bending of the magnetization causes the nonsingular vortex to decay into two singular $\SO(3)$ vortices in the FM phase, each described in this basis by $(e^{i\varphi},0,0)^\mathrm{T}$, taking $\varphi$ to be the azimuthal angle around each vortex line.

\begin{figure}[tb]
\centering
\includegraphics[width=\columnwidth]{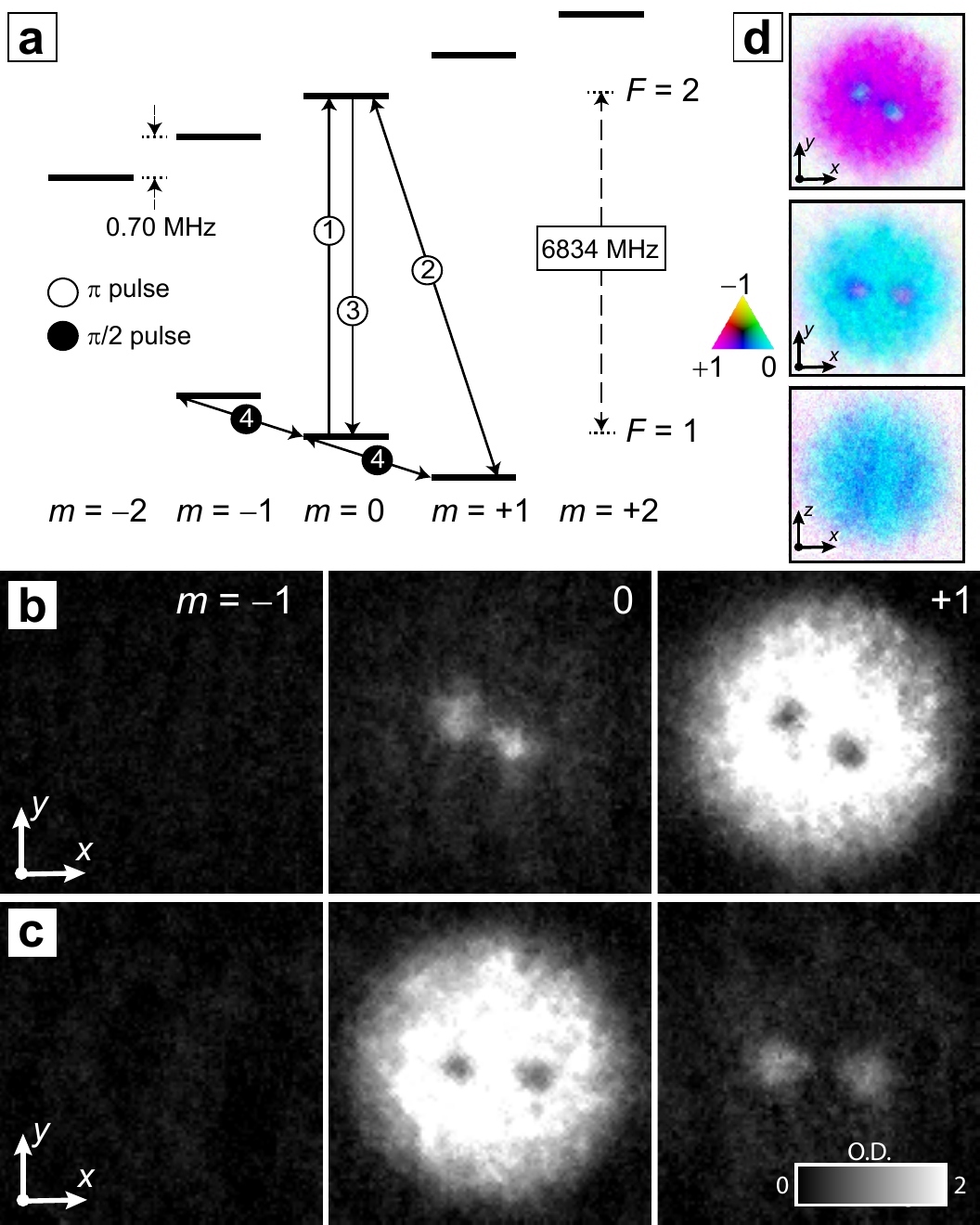}
\caption{Creation and detection of the polar SQVs. (a) A sequence of microwave $\pi$-pulses (open circles 1--3) swaps the $m=+1$ and $m=0$ components via the $F=2,m=0$ state, taking the initial pair of SO(3) vortices (b) into a pair of polar SQVs with filled FM cores (c). An optional radiofrequency $\pi/2$ pulse (filled circle 4) rotates the spinor. The respective pulse lengths are $(10,82,10,15)~\mu$s. (b,c) Top view of the spinor components before (b) and after (c) the phase exchange sequence, in units of optical depth (O.D.). (d) False color spinor composition, with a top image of the FM SQVs (upper) and top and side images of the polar SQVs (middle and bottom, respectively). The field of view in all panel images is $219~\micron \times 219~\micron$.\label{fig:polarvortex}}
\end{figure}

In scalar superfluids a singular defect implies that the superfluid density at the singularity vanishes, but in spinor BECs it is energetically favorable to accommodate the singularity by filling the vortex core with atoms in a different magnetic phase when the spin-dependent interaction is weaker than the spin-independent one~\cite{ruostekoski_prl_2003,lovegrove_pra_2012,kobayashi_pra_2012}. The vortex core regions in our experiment contain atoms in the nonrotating polar phase, described by $(0,1,0)^\mathrm{T}$, as shown in Fig.~\ref{fig:polarvortex}(b). These exhibit a coherent, stable topological interface between the two distinct magnetic phases~\cite{borgh_prl_2012,lovegrove_pra_2016}, where the magnetic phase changes continuously within the vortex core.  Analogous topological interfaces are universal across many areas of physics, ranging from superfluid liquid $^3$He~\cite{finne_rpp_2006,bradley_nphys_2008} to early-universe cosmology and superstring theory~\cite{kibble_jpa_1976,sarangi_plb_2002}, as well as to exotic superconductivity~\cite{bert_nphys_2011}.

The next step in the SQV creation process is the rapid exchange of the polar and FM phases. For the texture described above this amounts to swapping the $m=+1$ and $m=0$ spinor components with a sequence of three microwave pulses, as shown in Fig.~\ref{fig:polarvortex}(a). Afterwards, the topological interface between magnetic phases in the vortex core is reversed: along each singularity the atoms attain the pure FM phase, $(1,0,0)^\mathrm{T}$, while the surrounding circulating bulk superfluid is in the polar phase, $(0,e^{i\varphi},0)^\mathrm{T}$. These features are clearly seen in Fig.~\ref{fig:polarvortex}(c). Creation of this novel vortex state using the magnetic phase exchange technique is one of the principal results of the present study.

\begin{figure*}
  \centering
  \includegraphics[width=2\columnwidth]{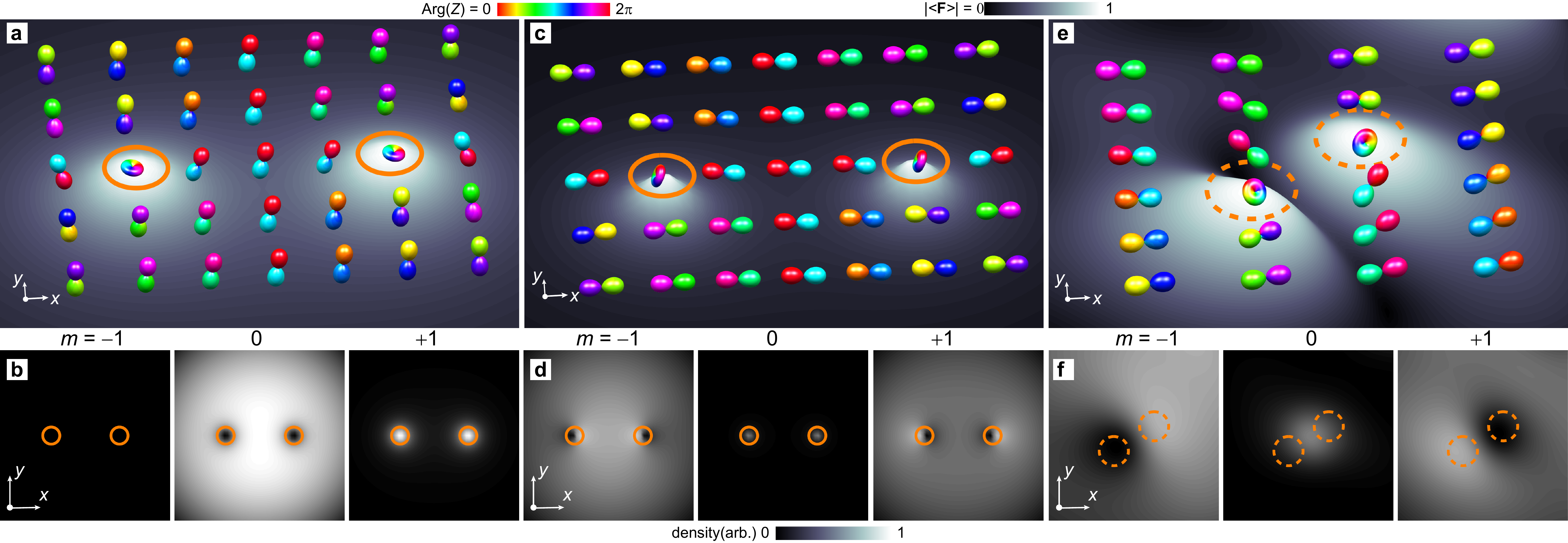}
  \caption{(a) Two polar SQVs corresponding to Fig.~\ref{fig:polarvortex}(c). The background indicates condensate spin magnitude, showing the FM cores. The symmetry of the order parameter is represented in terms of spherical harmonics \cite{supplemental}.
  (b) The corresponding spinor components.
  (c--d) Same as (a--b), but after application of a $\pi/2$ spinor rotation. (e--f) Expanded view of a single SQV after time evolution, corresponding to the experimental situation in Fig~\ref{fig:time-evolution}(a). The SQV has split into a pair of HQVs.
  Orange solid (dashed) circles indicate the positions of the SQVs (HQVs).
  }
  \label{fig:core-symmetry}
\end{figure*}

We model the two polar SQVs and their FM cores numerically in 3D [Fig.~\ref{fig:core-symmetry}(a)]. Each vortex in the figure represents a polar SQV with a filled core. Analytically, the spinor representing both the vortex and its FM core can be constructed as~\cite{lovegrove_pra_2016}
\begin{equation}
  \label{eq:general}
  \zeta = \frac{e^{i\varphi}}{2}
    \threevec{\sqrt{2}e^{-i\varphi} \left(D_-\sin^2\frac{\beta}{2}
      -D_+\cos^2\frac{\beta}{2}\right)}
     {-\left(D_- + D_+\right)\sin\beta}
     {\sqrt{2}e^{i\varphi}\left(D_-\cos^2\frac{\beta}{2}
       - D_+\sin^2\frac{\beta}{2}\right)},
\end{equation}
where $D_\pm=(1 \pm \absF)^{1/2}$ parameterizes the interpolation between the polar and FM phases as $\absF$ varies from $0$ in the bulk to $1$ on the vortex line. Here $\varphi$ again denotes the azimuthal angle around the vortex line, whereas $\beta$ is the polar angle that determines the order-parameter orientation, varying from $\beta=\pi/2$ away from the vortex line to $\beta=0$ on the line singularity itself. The initial state is modeled numerically and shown in Fig.~\ref{fig:core-symmetry}(a,b).

Although the spontaneous breaking of the defect core symmetry in the polar phase and the emergence of HQVs depend nontrivially on the relative interaction strength $c_2/c_0$~\cite{ruostekoski_prl_2003,Shinn2018,underwood_arxiv_2020}, this dependence can easily be obscured by the density gradients in a harmonically trapped BEC. Of additional significance is the effect of the applied magnetic field, which can restore the vortex core isotropy at sufficiently high $p$ and suppress the decay into HQVs at sufficiently high $q$~\cite{borgh_prl_2016a,underwood_arxiv_2020}. We find empirically that an applied bias magnetic field of 1~G is sufficient to inhibit evolution of the experimental SQV state depicted in Fig.~\ref{fig:polarvortex}(c) towards HQVs.

\begin{figure}[tb]
\centering
\includegraphics[width=\columnwidth]{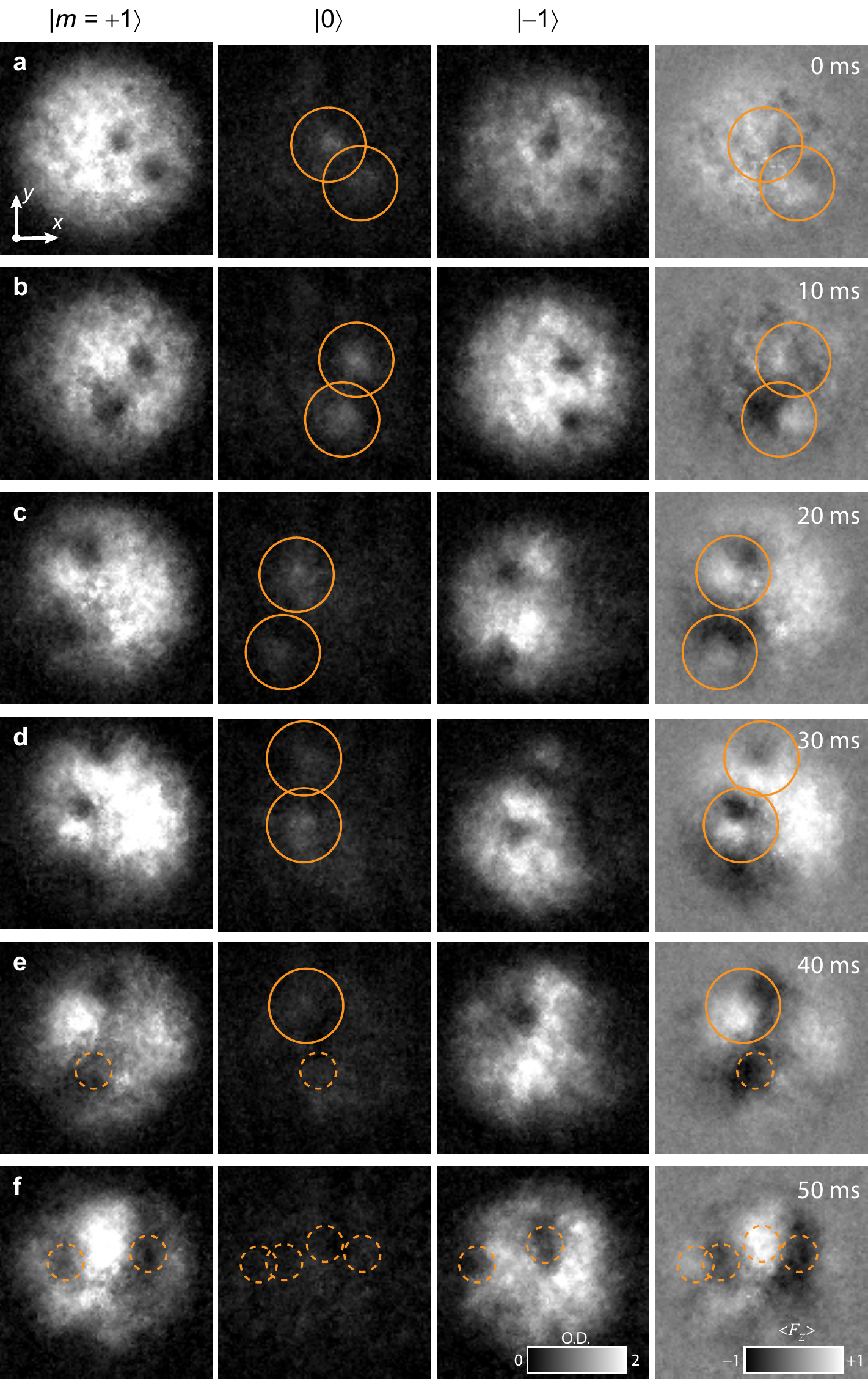}
\caption{The first 50~ms of time evolution after a the radio-frequency $\pi/2$-pulse. The first three columns show the atomic densities for the three spinor components; right column: longitudinal magnetization density ($n_{+1}-n_{-1}$). Solid circles indicate identifiable maxima in the $m=0$ spinor component and overlaid upon the composite image. Dashed circles in (e,f) are estimated locations of offset phase singularities identified from circumscribed density minima in the $m=\pm1$ spinor components. 
Each images has a field of view of $219~\micron \times 219~\micron$.\label{fig:time-evolution}}
\end{figure}

To induce condensate dynamics, we therefore apply a $\pi/2$-pulse within the $F=1$ manifold to rotate both the nematic director (in the polar phase) and the condensate spin (in the FM phase) into the $xy$ plane. The spinor rotation can be understood by describing a single SQV as a vortex line in the $m=0$ component with core filled by atoms in the $m=+1$ component.
Under a $\pi/2$ spinor rotation about the $y$ axis, the spinor transforms as
\begin{equation}
  \label{eq:spin-flip}
    \threevec {\sqrt{1-g(\rho)}}
             {e^{i\varphi}\sqrt{g(\rho)}}
             {0}
    \to
    \frac{1}{2}\threevec{-e^{i\varphi}\sqrt{2g(\rho)}+\sqrt{1-g(\rho)}}
                        {\sqrt{2-2(\rho)}}
                        {e^{i\varphi}\sqrt{2g(\rho)}+\sqrt{1-g(\rho)}}
\end{equation}
where $g(\rho) = \rho^2/(\rho^2+r_0^2)$ approximates the vortex-core profile with size parameterized by $r_0$. After the rotation, density maxima in the $m=0$ component appear at the locations of the vortex cores, each bracketed by symmetrically displaced density minima in the $m=\pm 1$ spinor components. These spatially offset phase singularities result from the sum of FM and polar terms of the initial spinor that transform differently within the vortex core~\cite{weiss_ncomm_2019,supplemental}. We experimentally observe these principal features immediately after the $\pi/2$ pulse [Fig.~\ref{fig:time-evolution}(a)].

As the system evolves, the spatial separation between the phase singularities associated with both SQVs increase, and each phase singularity grows in size as it fills with fluid in the $m=+1$ (or, for the other of the pair, $m=-1$) spinor component (Fig.~\ref{fig:time-evolution}).
These effects can be seen in the emergence of sharply defined proximate bright and dark regions in the total longitudinal magnetization density $M(\mathbf{r}) \approx n_{+1}-n_{-1}$~\cite{supplemental}, shown in the rightmost column of Fig.~\ref{fig:time-evolution}, as well as directly from the locations of the density minima within the $m=\pm 1$ spinor components.
The density of the $m=0$ spinor component also becomes more diffuse as the core region of each SQV grows.

We numerically simulate the dynamics using the coupled Gross--Pitaevskii equations derived from Eq.~\eqref{eq:hamiltonian}, also employing an algorithm to restore the conservation of longitudinal magnetization~\cite{lovegrove_prl_2014} in the presence of a small phenomenological dissipation. The resulting evolved state is shown in Fig.~\ref{fig:core-symmetry}(e--f) where the pair of phase singularities associated with each SQV has formed a split pair of HQVs. In the simulation, the fully separated FM cores and the symmetry representation of the wavefunction shown in Fig.~\ref{fig:core-symmetry}(e--f) permit straightforward identification of the HQVs. Once separated, the characteristic size of the filled regions is theoretically established by the spin healing length~\cite{lovegrove_pra_2012,ruostekoski_prl_2003}, which is much larger than the density healing length.

Connecting the experimentally obtained spinor component densities to the corresponding vortex states in the simulation requires some care, especially with respect to the presence of the $m=0$ spinor component. For the given magnetic field direction and the idealized case of a condensate with polar interactions in the easy-plane polar regime~\cite{supplemental}, an empty-core SQV splits into two spatially offset phase singularities in the $m=\pm 1$ spinor components and the $m=0$ component remains absent. These phase singularities are unambiguously HQVs, each surrounded by polar fluid where $\nematic$ remains in the plane. For FM interactions in the EAP regime, however, the presence of the phase singularities in regions where the $m=0$ component is nonzero can indicate either the existence of nonzero transverse spin within the unsplit SQV core [Fig.~\ref{fig:core-symmetry}(c--d)], or the rotation of $\nematic$ out of the $xy$ plane in a fully split pair of HQVs [Fig.~\ref{fig:core-symmetry}(e--f)]. Only the disappearance of the $m=0$ spinor component in the experimental images, implying fully longitudinal spin domains with a director that remains in the $xy$ plane, conclusively announces the presence of two HQV. This is approximately the situation in Fig.~\ref{fig:time-evolution}(e--f).

There are several effects that conspire to complicate the experimental interpretation of our results. 
First, the offset phase singularities of the two initial singular vortices may closely approach one another, making their disambiguation problematic. The approximate SQV locations may often be located by identifying density maxima in the $m=0$ spinor component, as shown in Fig.~\ref{fig:time-evolution}(a--d). The presence of a small and consistent but uncontrolled magnetic field gradient globally shifts the $m=\pm 1$ spinor components with respect to one another, creating less sharply defined regions of opposite magnetization on a size scale comparable to that of the condensate. This effect is most pronounced in Fig.~\ref{fig:time-evolution}(c,d), where the $m=+1$ ($m=-1$) spinor component is shifted to the right (left). Nevertheless, the vortices may often be located by identifying the density minima associated with the offset phase singularities in the $m= \pm 1$ spinor components, as suggested by the dashed circles in Fig.~\ref{fig:time-evolution}(e) and~(f). The singularities can still be difficult to discern if they are near the edge of the condensate, if they are tilted with respect to the imaging axis, or if they exhibit longitudinal (Kelvin wave) excitations~\cite{bretin_prl_2003}. One of the expected singularities in Fig.~\ref{fig:time-evolution}(e) likely cannot be cleanly identified as a result of one or more of these 3D effects.

We have implemented a controllable technique of rapid magnetic phase exchange to facilitate the controlled creation of a pair of singular SQVs with nonrotating FM cores in the polar magnetic phase of a spin-1 superfluid with FM interatomic interactions. Our experimental and theoretical analysis of the decay process in three dimensions suggests the emergence of HQVs. Similar techniques may be used to generate pairs of filled-core vortices in the magnetic phases of spin-2 condensates, where vortex collisions are predicted to possess a non-Abelian character~\cite{kobayashi_prl_2009,borgh_prl_2016,semenoff_prl_2007,mawson_prl_2019} and for which the topological interfaces may lead to exotic phenomena such as vortices with triangular cores~\cite{borgh_prl_2016}.

\begin{acknowledgments}
We gratefully acknowledge experimental assistance and helpful conversations with T. Ollikainen. D.S.H.\ acknowledges financial support from  the National Science Foundation (Grant No.\ PHY--1806318.) and J.R.\ from the UK EPSRC (Grant Nos.\ EP/P026133/1, EP/M013294/1).
\end{acknowledgments}


%

\widetext
\clearpage

\setcounter{equation}{0}
\setcounter{figure}{0}
\setcounter{section}{0}
\renewcommand{\theequation}{S-\arabic{equation}}
\renewcommand{\thefigure}{S-\arabic{figure}}

\section*{Supplemental Material for ``Controlled Creation and Decay of Singly-Quantized Vortices in a Polar Magnetic Phase''}

\section{Magnetic Phases}

The two distinct ground-state phases in a spin-1 Bose--Einstein condensate at zero field are determined by the sign of the spin-dependent atomic interaction parameter $c_2$ in Eq.~(1). For $c_2<0$ (e.g., for $^{87}$Rb) the system energy is maximized for $\absF=1$, resulting in a ferromagnetic (FM) order-parameter space characterized by the group of 3D rotations, $\SO(3)$~\cite{ho_prl_1998,ohmi_jpsj_1998}. Contrariwise, for $c_2>0$ (e.g., for $^{23}$Na) the system energy is minimized for $\absF=0$, resulting in a uniaxial nematic (polar) order-parameter space characterized by $[S^2 \times \mathrm{U(1)}]/{\mathbb{Z}_2}$ with local $\U(1)$ phase $\tau$ and nematic axis $\nematic$~\cite{leonhardt_jetplett_2000,zhou_prl_2001,zhou_ijmpb_2003}. Here, the two-element factor group $\mathbb{Z}_2$ appears due to the symmetry $\nematic e^{i\tau} = -\nematic e^{i(\tau+\pi)}$. It is this symmetry that permits vortices in the polar phase to carry half-integer circulation when $\tau$ runs between $0$ and $\pi$ around the vortex singularity and $\nematic$ concurrently rotates by $\pi$.

At stronger fields with fixed magnetization the quadratic Zeeman term $q$ becomes important for determining the ground state~\cite{zhang_njp_2003,murata_pra_2007,sadler_nature_2006,ruostekoski_pra_2007}, and within the polar phase itself there arise two relevant phases: the easy-axis polar (EAP), where $\nematic$ is aligned with an applied magnetic field, and the easy-plane polar (EPP), where $\nematic$ is perpendicular to the applied magnetic field. In our experiment $q > |c_2| n$, specifying an EAP ground-state phase.

\begin{figure}[hb]
\centering
\includegraphics[width=0.75\columnwidth]{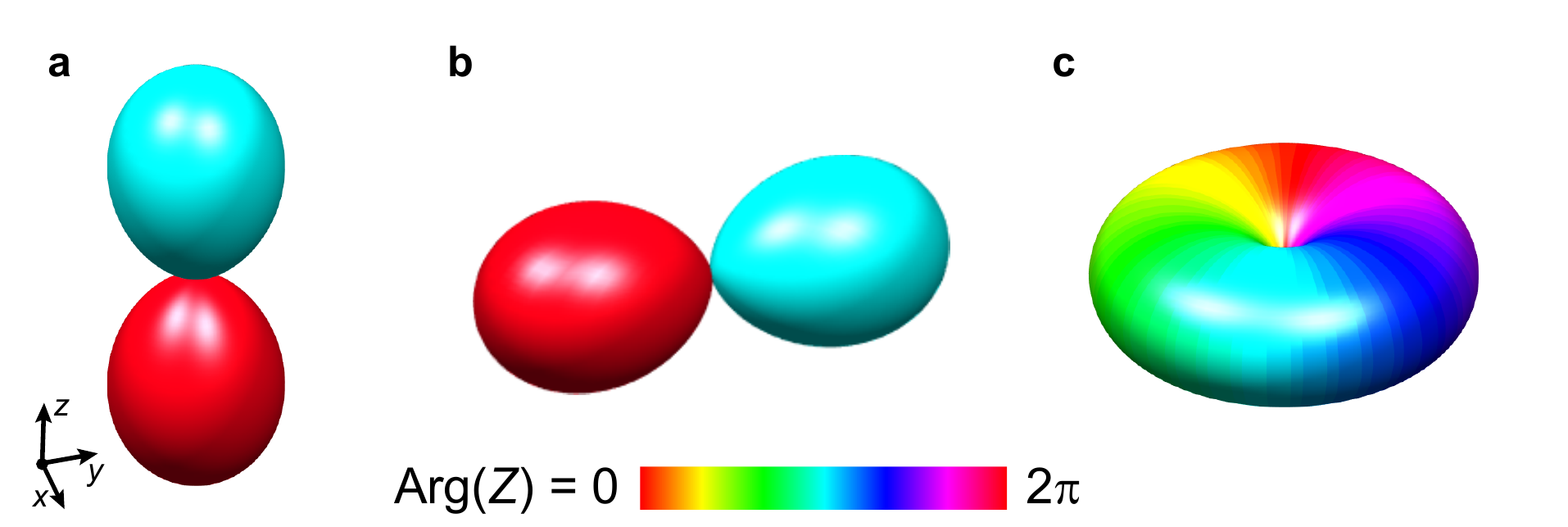}
\caption{Symmetries of the ground-state magnetic phases of a spin-1 Bose--Einstein condensate, expressed as the surface of $|Z(\theta,\phi)|^2$ with color representing $\mathrm{Arg}(Z)$. A magnetic field is applied along the $z$ axis. (a) The EAP phase, with representative spinor $(0,1,0)^\mathrm{T}$. (b) The EPP phase, with representative spinor $(1,0,1)^\mathrm{T}/\sqrt{2}$. (b) The FM phase, with representative spinor $(1,0,0)^\mathrm{T}$.\label{fig:sphhrm}}
\end{figure}

The order parameter symmetry in Fig.~2 is represented by the surface of $|Z(\theta,\phi)|^2$, where $Z(\theta,\phi)=\sum_{m=-1}^{+1}Y_{1,m}(\theta,\phi)\zeta_m$ expands the spinor in terms of the spherical harmonics $Y_{1,m}(\theta,\phi)$, with local spherical coordinates $(\theta,\phi)$, and gauge color given by $\mathrm{Arg}(Z)$. Simple examples of the polar and FM ground-state phases are shown in Fig.~\ref{fig:sphhrm}.

\clearpage

\section{Vortex Lines in Three Dimensions}

\begin{figure}[hb]
\centering
\includegraphics[width=0.75\columnwidth]{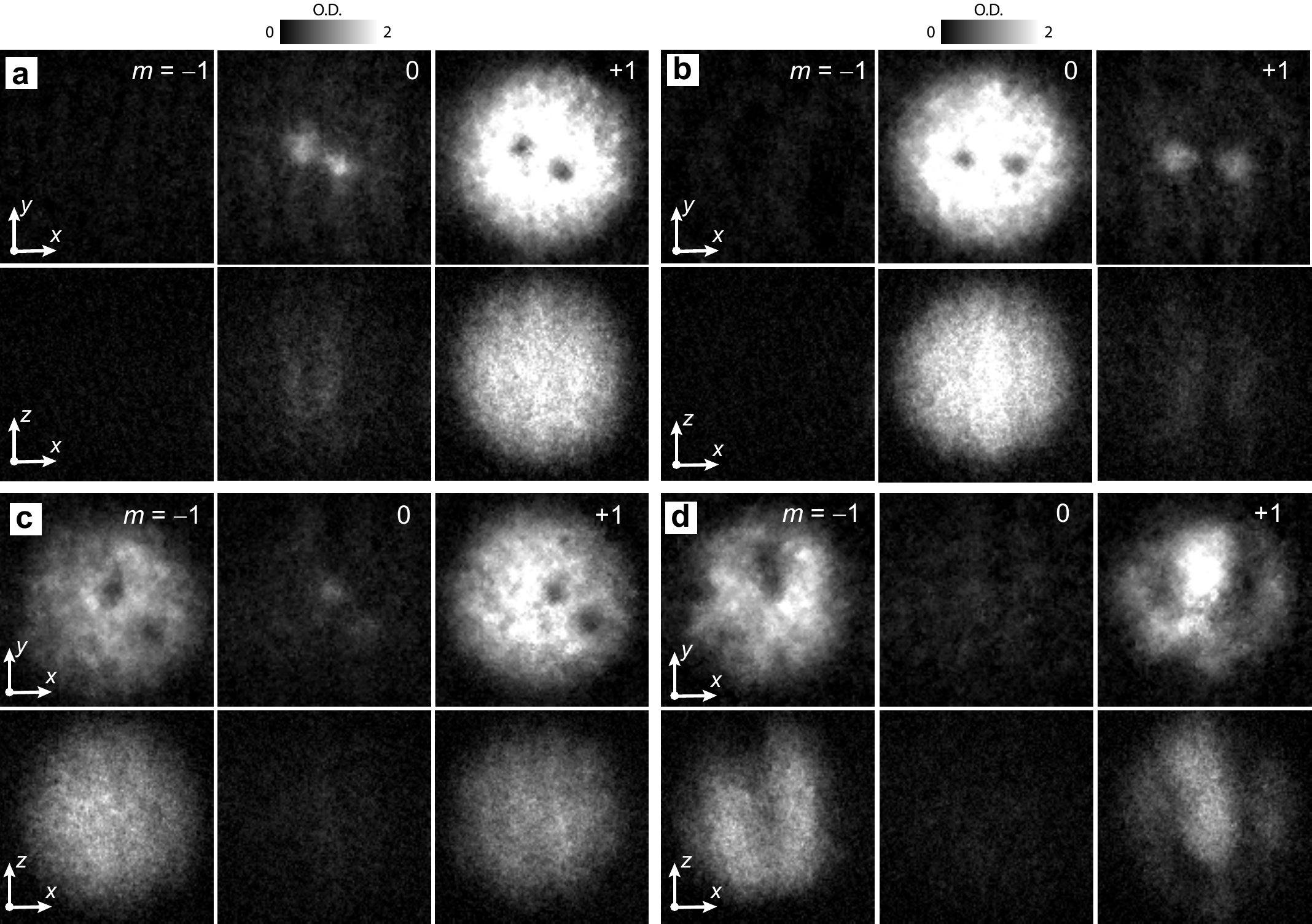}
\caption{Top and side images from several representative condensates. (a) The initial $\SO(3)$ vortices, after the splitting of the nonsingular texture. (b) The pair of polar, singly-quantized vortices (SQVs) after the rapid exchange of the magnetic phases. (c) The two SQVs after rotating the directors into the $xy$ plane by the radio-frequency $\pi/2$-pulse. (d) One SQV after 50~ms of time evolution. The intensity in all figures is in terms of optical depth (O.D.), and the field of view is $219~\micron \times 219~\micron$.\label{fig:vortlines}}
\end{figure}

The Bose--Einstein condensate in our experiment is mildly oblate, as it is confined in a harmonic potential with trap frequencies $(\omega_r,\omega_z) = 2\pi(130,170)~\mathrm{s}^{-1}$. As a result, the vortices are lines in three dimensions rather than the points associated with highly oblate, quasi two-dimensional condensates. Several representative images of the expanded condensates, taken simultaneously from both the top and side, are shown in Fig.~\ref{fig:vortlines}.

\vfill
\pagebreak

\section{Additional Numerical Simulations}

We illustrate here some of the complications that arise in the interpretation of the experimental SQV evolution images with the help of Gross--Pitaevskii (GP) simulations of energy relaxation and dynamics of a single SQV. The GP equations are derived from Eq.~(1) of the main text, in which the trap parameters are also defined. The simulations integrate the GP equations using the split-step method~\cite{javanainen_jpa_2006} and an algorithm to conserve the longitudinal magnetization~\cite{lovegrove_prl_2014}.

\begin{figure}[ht]
  \centering
  \includegraphics[width=\linewidth]{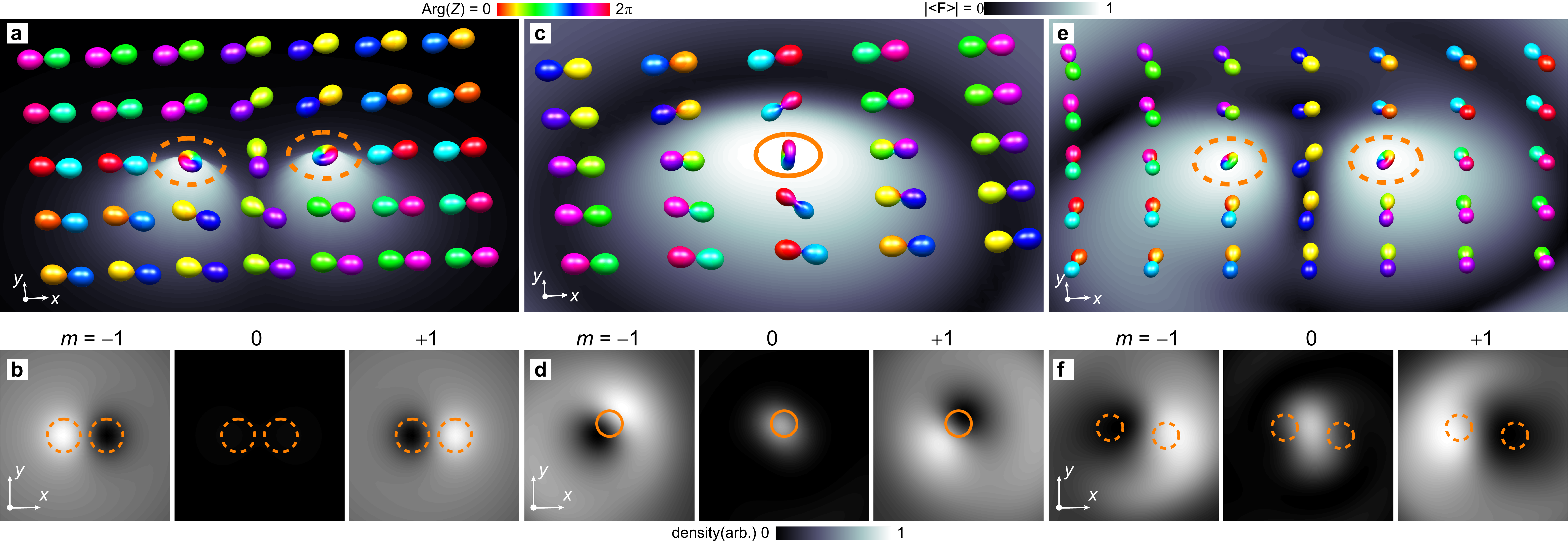}
  \caption{Numerical simulation of SQV core deformation and HQV formation. (a) Spin magnitude (background surface) and order-parameter symmetry in the spherical-harmonics representation after pure energy relaxation in a $^{23}$Na condensate $c_0/c_2 \simeq 28$) with $q=-0.1\hbar\omega_r$, resulting in HQVs. (b) Corresponding spinor-component densities. (c--d) As (a--b) for a $^{87}$Rb condensate ($c_0/c_2 \simeq -216$) with $q=0.01\hbar\omega_r$ and $\eta=0.02$. The spinor components exhibit density profiles similar to HQVs even though the vortex has not split. (e--f) As (c--d), but with complete splitting for $q=0.55\hbar\omega_r$, corresponding to the experiment, and $\eta=0.008$.
  \label{fig:interpretation}}
\end{figure}

Pure energy relaxation (i.e., propagation of the GP equations in imaginary time) for a single SQV in a $^{23}$Na BEC in the EPP regime provides the reference scenario [Fig.~\ref{fig:interpretation}(a)]. The vortex in this case is known to relax into a pair of HQVs~\cite{lovegrove_pra_2012}. The HQVs are readily identified in the simulation by the order-parameter symmetry, also showing continuous deformation to the FM vortex core. The vector $\nematic$ remains everywhere in the $xy$ plane, yielding a completely depopulated $m=0$ spinor component and equal bulk densities in the $m=\pm1$ components. The vortices appear in the form first proposed by Leonhardt and Volovik~\cite{leonhardt_jetplett_2000}, as offset phase singularities in the populated components. In this case the HQVs can conversely be inferred directly from the component images where the $m=0$ spinor component vanishes, as this corresponds directly to the $z$-component of $\nematic$.

As noted in the main text, however, the presence of the $m=0$ spinor component in the experimental images does not determine whether an SQV has split into two HQVs. By way of further numerical illustration we consider the time evolution of an SQV prepared initially in the EPP phase, but in a $^{87}$Rb BEC with $q>0$ (FM atom-atom interactions). Dissipation, included here phenomenologically by taking $t\to(1-i\eta)t$ (where $\eta \ll 1$) in the GP equations, causes the condensate to evolve towards the FM phase; on the other hand, the Zeeman energy favors rotating $\nematic$ out of the $xy$ plane. Both effects may lead to a nonzero density remaining in the $m=0$ component, which can correspond to \emph{either} an unsplit SQV [Fig.~\ref{fig:interpretation}(b)] or two separated HQVs, where $\nematic$ rotates out of the $xy$ plane between the vortex lines [Fig.~\ref{fig:interpretation}(c)].

\pagebreak

\begin{figure}[ht]
  \centering
  \includegraphics[width=\linewidth]{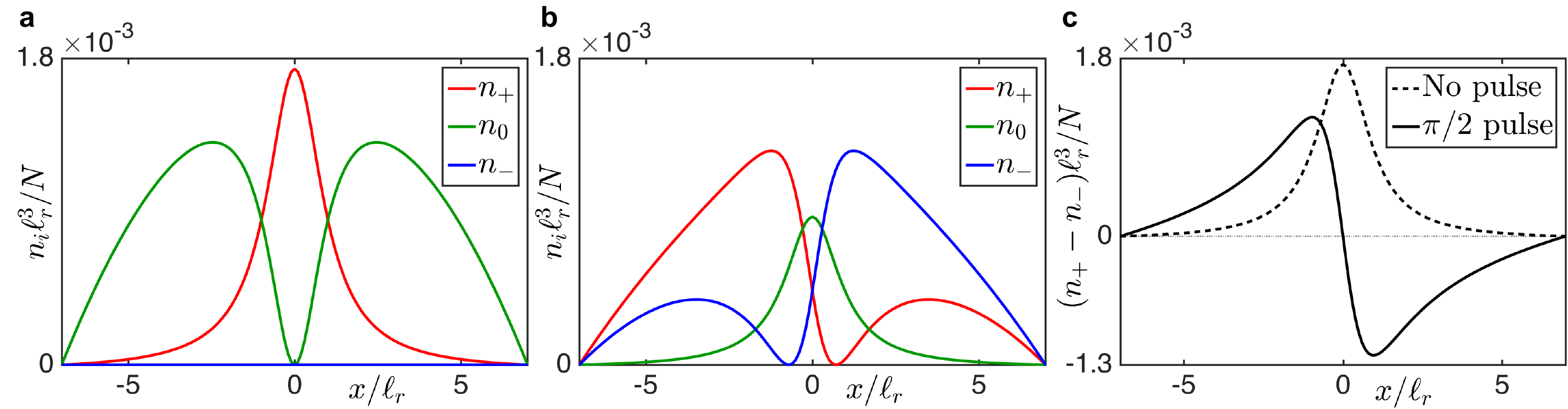}
  \caption{Analytically constructed cross-sections of spinor component densities and magnetization densities of an SQV. (a) The spinor components for the initial SQV in the EAP phase. (b) The spinor components for the SQV after a $\pi/2$ spin rotation to the EPP phase. Offset phase singularities appear near the center of the plot, where the $m=\pm 1$ components vanish but the $m=0$ component does not. (c) Density difference $n_+-n_-$, to which the longitudinal magnetization density is proportional, before and after the $\pi/2$-pulse. The strongly magnetized regions coincide with the phase singularities in (b). Lengths given in units of the transversal trap length $\ell_r=\sqrt{\hbar/(M\omega_r)}$, for atomic mass $M$, and $N$ is the number of atoms.
  \label{fig:lineplots}}
\end{figure}

We further illustrate the experimental signatures by analytically modeling the effect of the $\pi/2$-pulse applied to rotate the BEC from the EAP into the EPP phase immediately after initial preparation of the SQVs. We assume that the condensate away from the vortex core is exactly in these phases before and after the $\pi/2$-pulse, respectively, which is a good approximation of the experimental situation. Figure~\ref{fig:lineplots} corresponds directly to Eq.~(5) in the main text, showing the spinor component density profiles $\{n_0,n_{\pm}\}$ for a single SQV before and after the $\pi/2$-pulse. The spin rotation itself results in offset phase singularities in the $m=\pm1$ components with non-zero $n_0(\rr)$ at the vortex singularity between them, illustrating that the presence of the offset phase singularities alone does not imply splitting of the SQV. This effect is seen in the experiment in Fig.~3(a)--(d) in the main text. The $\pi/2$ pulse also rotates the condensate spin inside the vortex core resulting in offset, oppositely polarized peaks in the longitudinal magnetization density, presented as $n_+(\rr)-n_-(\rr)$ in Fig.~\ref{fig:lineplots}(c), that coincide with the phase singularities. These peaks thus appear before any splitting of the SQV has taken place. After the splitting the peaks locate the HQV cores, but are not in themselves complete evidence of the splitting.

\end{document}